\documentclass[12pt]{article}

\usepackage{amsmath,amsfonts,latexsym,amssymb,amscd}
\usepackage{pslatex}
\usepackage[latin1]{inputenc}
\usepackage[T1]{fontenc}
\usepackage{pspicture}
\usepackage{verbatim,amsthm,curves,graphics}
\usepackage{mathrsfs}

\usepackage{graphicx}

\newcommand{\mr}{{\mathbb R}}

\begin{document}


\title{ Classification of static asymptotically flat spacetimes with a photon sphere in Einstein-multiple-scalar field theory  }

\author{
Stoytcho Yazadjiev$^{}$\thanks{\tt yazad@phys.uni-sofia.bg}
\\ \\
{\it $ ^1$Department of Theoretical Physics, Faculty of Physics, Sofia
University} \\
{\it 5 J. Bourchier Blvd., Sofia 1164, Bulgaria} \\
{\it $ ^2$Theoretical Astrophysics, Eberhard-Karls University of T\"ubingen,}\\ {\it T\"ubingen 72076, Germany  }  \\
{\it Institute of Mathematics and Informatics, 	Bulgarian Academy of Sciences,}  \\ {  	Acad. G. Bonchev St. 8, Sofia 1113, Bulgaria} }

\date{}

\maketitle

\begin{abstract}In the present paper we consider the Einstein-multiple-scalar field theory. When the target space of the scalar fields is a
complete, simply connected Riemannian manifold with non-positive sectional curvature we prove that the static and asymptotically flat solutions
which posses a photon sphere are uniquely specified by their asymptotic data, i.e. by their   mass, scalar charges and asymptotic values of the scalar fields. The unique solution with a photon sphere and prescribed asymptotic data is explicitly constructed.

\end{abstract}


\sloppy

\section{Introduction}

The geometrical photon structures like the photon rings, the photon spheres and the photon domains play a crucial role in  the strong gravitational lensing and the formation  of the shadow of the  compact objects \cite{Virbhadra2002}-\cite{Perlick_2021}. That is why the geometrical photon structures and their properties have been intensively studied from both physical and mathematical point of view \cite{Claudel2001}-\cite{Kobialko_2021}. The photon spheres are of  particular interest due to their special properties. In some sense the photon spheres resemble the black hole horizons. As the presence of a horizon allows us to classify the spacetimes in terms of their conserved asymptotic quantities, so the presence of a photon sphere also  allows us to classify some  static spacetimes only in terms of their asymptotic charges \cite{Cederbaum2014}-\cite{Jahns_2019}.

The present paper aims to prove a classification theorem for the static and asymptotically 
flat solutions of the Einstein-multiple-scalar field equations, containing a photon sphere. This comes as a logical continuation of our previous work \cite{Yazadjiev2015} which proved similar theorem in the case of a single scalar field. The Einstein-multiple-scalar field equations naturally arise in the context of the higher dimensional theories under dimensional reduction as well they can be considered as the vacuum sector of the tensor-multi-scalar theories of gravity. Also, the Einstein-Maxwell and the Einstein-Maxwell-dilaton (-axion) gravity with a static symmetry effectively reduce to the Einstein-multiple-scalar field equations and therefore our consideration here cover our previous theorems proven in \cite{Yazadjiev2015a,Yazadjiev2015b} as particular cases.

We prove that when the target space  of the scalar fields is a  complete, simply connected Riemannian manifold with a 5non-positive scalar curvature, the static and  asymptotically flat solutions to the  Einstein-multiple-scalar field equations possessing a photon sphere are uniquely specified by their mass $M$, scalar charges $q^a$ and the asymptotic values of the scalar fields $\varphi^a_\infty$.

\section{General definitions and equations}

In the present paper we consider the Einstein-multiple-scalar field theory. More precisely, let $(M, g^{(4)})$ is the 4-dimensional spacetime and $({\cal E}_N,\gamma)$ is a $N$-dimensional Riemannian manifold with metric $\gamma$ (the so-called  target space) . We consider a map $\varphi: (M,g)\to ({\cal E}_N,\gamma)$  and its deferential $d\varphi$ induces a map between the tangent spaces of $M$ and ${\cal E}_N$, $d\varphi: TM\to T{\cal E}_N$. The norm of the differential will be denoted by $<d\varphi,d\varphi>$. In local coordinate patches on 
$M$ and ${\cal E}_N$ we have 

\begin{eqnarray}
<d\varphi,d\varphi>= g^{(4)\,\mu\nu}(x)\gamma_{ab}(\varphi(x))\partial_\mu\varphi^a(x)\partial_\nu\varphi^b(x).
\end{eqnarray} 

The Einstein-multiple-scalar  field theory is  given by the action

\begin{eqnarray}\label{action}
S=\frac{1}{16\pi G} \int_{M} d^4x\sqrt{-g}\left[R^{(4)} - 2<d\varphi,d\varphi> \right],
\end{eqnarray}
where  $R^{(4)}$ is the Ricci scalar curvature. The field equations corresponding to the action (\ref{action}) and written in local coordinate patches of  $M$ and ${\cal E}_N$ are the following 

\begin{eqnarray}\label{FE1}
&&R^{(4)}_{\mu\nu} = 2\gamma_{ab}(\varphi)\nabla^{(4)}_{\mu}\varphi^a \nabla^{(4)}_{\nu}\varphi^b, \\
&&\nabla^{(4)}_{\mu}\nabla^{(4)\mu}\varphi^a= - \gamma^{a}_{cd}(\varphi)\nabla^{(4)}_{\mu}\varphi^c \nabla^{(4)\mu}\varphi^d , 
\end{eqnarray}
where $\nabla^{(4)}_{\mu}$ is the covariant derivative with respect to the spacetime metric $g^{(4)}_{\mu\nu}$ and 
$\gamma^{a}_{cd}(\varphi)$ are the Christoffel symbols with respect to the target space metric $\gamma_{ab}(\varphi)$.

In the present work we are interested in static and asymptotically flat spacetimes. A spacetime is called static if there exists a smooth Riemannian manifold $(M^{(3)}, g^{(3)}_{ij})$ and a smooth lapse function $N: M^{(3)} \to {\mr}^{+}$ such that

\begin{eqnarray}
M^{(4)}= {\mr} \times M^{(3)}, \,\,\, g^{(4)}_{\mu\nu}dx^\mu dx^\nu= - N^2 dt^2 + g^{(3)}_{ij}dx^idx^j.
\end{eqnarray}
The scalar fields are called static if ${\cal L}_{\xi}\varphi^a= 0$ where ${\cal L}_{\xi}$ is
the Lie derivative along the Killing field $\xi=\frac{\partial}{\partial t}$. Both  notions of staticity are consistent since the Ricci
1-form $Ric^{(4)}[\xi]=\xi^{\mu}Ric^{(4)}_{\,\mu\nu}\,dx^{\nu}$ is zero due to the field equations and the fact that $\xi^{\mu}\nabla^{(4)}_{\mu}\varphi^a={\cal L}_{\xi}\varphi^a= 0$.

We  adopt the following notion of asymptotic flatness. A spacetime is called asymptotically flat if there exists a compact set $K \subset M^{(3)}$ such that
$M^{(3)} - K$ is diffeomorphic to ${\mr}^{3} \backslash \bar{B}$ where $\bar{B}$ is the closed unit ball centered at the origin in ${\mr}^3$, and such that

\begin{eqnarray}
g^{(3)}_{ij}= \delta_{ij} + {\cal O}(r^{-1}), \;\;\; N= 1 - \frac{M}{r} + {\cal O}(r^{-2}), \;\; \varphi^a =\varphi^a_{\infty} - \frac{q^a}{r} + {\cal O}(r^{-2}),
\end{eqnarray}
with respect to the standard radial coordinate $r$ of $\mr^3$. Here $M$, $\varphi^a_{\infty}$ and $q^a$ are constants with $M$ and $q^a$ being the mass and the scalar charges, respectively. We will consider spacetimes with $M>0$. It is also convenient to introduce the {\it generalized scalar charge} $Q$
defined by 

\begin{eqnarray}\label{GSC}
Q^2 = \gamma_{ab}(\varphi_\infty)q^a q^b. 
\end{eqnarray}

The dimensionally reduced static Einstein-multiple-scalar field equations are the following

\begin{eqnarray}\label{DRFE}
&&Ric^{(3)}_{ij} = N^{-1} \nabla^{(3)}_i \nabla^{(3)}_j N + 2 \gamma_{ab}(\varphi)\nabla^{(3)}_i\varphi^a \nabla^{(3)}_j\varphi^b ,\nonumber \\
&&\nabla^{(3)}_i \nabla^{(3)i}N=\Delta^{(3)}N=0, \\
&&\nabla^{(3)}_i\left(N \nabla^{(3)i}\varphi^a\right)=-N \gamma^{a}_{bc} \nabla^{(3)i}\varphi^a \nabla^{(3)}_i \varphi^b , \nonumber
\end{eqnarray}
where $\nabla^{(3)}_i$ and $Ric^{(3)}_{ij}$ are the Levi-Civita connection and the Ricci tensor with respect to the metric $g^{(3)}_{ij}$.

Now we can define the notion of a photon sphere. First we give the definition of a  photon surface \cite{Claudel2001,Cederbaum2014}.

\medskip
\noindent

{\bf Definition} {\it  An embedded timelike hypersurface ${\cal P}\hookrightarrow M^{(4)}$  is called a photon surface if and only if any null geodesic initially tangent to
${\cal P}$ remains tangent to ${\cal P}$ as long as it exists. }

\medskip
\noindent

For the  photon sphere we adopt the following definition which is an extension of the definition in the case with a single scalar field, namely  

\medskip
\noindent

{\bf Definition} {\it Let ${\cal P}\hookrightarrow M^{(4)}$ be  a photon surface. Then  ${\cal P}$  is called a photon sphere if the lapse function $N$ and the scalar fields $\varphi^a$  are constant along ${\cal P}$. }

\medskip
\noindent

As a further technical assumption we shall assume that the lapse function regularly foliates the region of  spacetime $M^{(3)}_{ext}$ exterior to the photon sphere which means that $g^{(3)}(\nabla^{(3)}N,\nabla^{(3)}N)\ne 0$  everywhere on $M^{(3)}_{ext}$. We also define the function  $\rho : M^{(3)}_{ext} \to \mr^{+} $ by

\begin{eqnarray}
\rho=\left[g^{(3)}(\nabla^{(3)}N,\nabla^{(3)}N)\right]^{-1/2}.
\end{eqnarray}

The 2-dimensional intersection  of the photon sphere ${\cal P}$ and the time slice $M^{(3)}$ will be denoted by $\Sigma$. By the very definition of the photon sphere,   $\Sigma$ which is the inner boundary of $M^{(3)}_{ext}$, is given by $N=N_{0}$ for some $N_{0}\in \mr^{+}$. The metric induced on $\Sigma$ will be denoted by $\sigma$. We also note  that our assumptions restrict our considerations to the case of a connected photon sphere. All the level sets $N=const$, including $\Sigma$, are topological spheres which is a direct consequence from our assumptions.

By the maximum principle for harmonic functions and by the asymptotic behavior of $N$ for $r\to\infty$ we obtain that the values of $N$ on $M_{ext}^{(3)}$ satisfy $N_{0}\le N<1 $.

\section{Scalar map as a geodesic map for  target spaces with non-positive curvature}

Here we will focus on target spaces which are complete, simply connected Riemannian manifolds with a non-positive sectional curvature. The last condition  means that the curvature tensor of the target space ${\cal R}_{abcd}$ satisfies

\begin{eqnarray}
{\cal R}_{abcd}W^{ab}W^{cd}\le 0
\end{eqnarray}
for any $W^{ab}$.

In this section it is more convenient to work with the 3-metric $h_{ij}= N^2  g^{(3)}_{ij}$. In terms of $h_{ij}$ the dimensionally reduced
field equations take the form 

 \begin{eqnarray}\label{DRFEH}
&&R(h)_{ij}= 2D_{i}uD_{j}u + 2 \gamma_{ab}(\varphi)D_{i}\varphi^a D_{j}\varphi^a ,\nonumber\\
&&D_{i}D^{i} u = 0, \\
&&D_{i}D^{i}\varphi^a = -\gamma^a_{bc}(\varphi) D_i\varphi^b D^i\varphi^c, \nonumber
\end{eqnarray}
where $u=\ln(N)$ and  $D_{i}$ and $R(h)_{ij}$ are the Levi-Chivita connection and the Ricci tensor with respect to $h_{ij}$, respectively. 

Now we formulate the key result on which the classification  is based. 

\medskip
\noindent

{\bf Theorem} {\it Let the target space $({\cal E}_N,\gamma_{ab})$ be a complete, simply connected Riemannian manifold with non-positive sectional curvature. Then, for fixed $(M,\varphi^a_\infty, q^a)$, the scalar map $\varphi$ is a geodesic map. In other words,	the scalar fields $\varphi^a$ and the lapse function $N$ are subject to the functional dependence  $\varphi^a=\psi^a(u)$ where $\psi^a(u)$ is the unique affinely parameterized geodesic on ${\cal E}_N$ determined by the conditions $\psi^a(0)=\varphi^a_{\infty}$ and $\frac{d\psi^a}{du}(0)=\frac{q^a}{M}$. }

\medskip
\noindent

{\bf Proof:} By the very definition of the photon sphere we have that the restrictions of $u$ and $\varphi^a$ on  $\Sigma$ are constants and these constants will be denoted by $u_0$  and $\varphi^a_0$, respectively. Since $({\cal E}_N,\gamma_{ab})$ is complete, simply connected with non-positive sectional curvature there exists a unique geodesic $\psi :I \to {\cal E}_N$  connecting the points $\varphi_0=\{\varphi^a_0\}$ and $\varphi_{\infty}=\{\varphi^a_{\infty}\}$. Without loss of generality we can chose
the affine parameter of the geodesic $\psi$ to vary in the interval $I=[u_0,0]$ with $\psi(u_0)=\varphi_0$ and $\psi(0)=\varphi_\infty$.    
As a next step we construct the map ${\tilde \varphi}$ defined by ${\tilde \varphi}=\psi \circ u : M^{(3)}_{ext} \to {\cal E}_N$ (${\tilde \varphi}^a=\psi^a(u)$).
This map satisfies the equations for the scalar fields. Indeed we have 

\begin{eqnarray}
D_{i}D^{i}{\tilde\varphi}^a  +\gamma^a_{bc}({\tilde \varphi}) D_i{\tilde\varphi}^b D^i{\tilde \varphi}^c= 
\left[\frac{d^2\psi^a}{du^2} + \gamma^{a}_{bc}(\psi(u)) \frac{d\psi^b}{du}\frac{d\psi^c}{du} \right] D_iu D^iu + \frac{d\psi^a}{du} D_iD^i u=0 
\end{eqnarray}
since $\psi$ is affinely parameterized geodesic on ${\cal E}_N$ and $D_iD^i u=0$. 
Will call the maps ${\tilde \varphi}$ {\it geodesic maps}. Now we shall show that  $\varphi$ is a geodesic map. This can be done in the following way. Consider the unique geodesic  ${\cal T}$  on ${\cal E}_N$ connecting $\varphi$ and ${\tilde \varphi}$ with an affine parameter $0\le \tau \le1$. Then the geodesic distance $S(\varphi,{\tilde \varphi})$ between the  $\varphi$ and ${\tilde \varphi}$ satisfies the Bunting 
identity \cite{Bunting_1983,Carter_1985}

\begin{eqnarray}
D_{i}D^{i} S^2 = \int_0^1 d\tau \left( {\hat \nabla}^i s_a {\hat \nabla}_i s^a -{\cal R}_{abcd}s^a{\hat \nabla}_i{\cal T}^b s^c{\hat \nabla}^i{\cal T}^d  \right)
\end{eqnarray} 
where $s^a=\frac{d{\cal T}^a}{d\tau}$ and ${\hat \nabla}_is^a= \partial_i{\cal T}^b \nabla_{b}s^a=\partial_is^a + \partial_i{\cal T}^c  \gamma^a_{cb}s^b$. Taking into account that the sectional curvature of ${\cal E}_N$ is non-positive we obtain that $D_{i}D^{i} S^2\ge 0$, i.e.
$S(\varphi,{\tilde \varphi})$ is a subharmonic function on $M^{(3)}_{ext}$. Since $S(\varphi,{\tilde \varphi})=0$  on $\partial M^{(3)}_{ext}=\Sigma$ (i.e. $\varphi|_\Sigma = {\tilde \varphi}|_\Sigma=\varphi_0$) we conclude that $S(\varphi,{\tilde \varphi})=0$ everywhere 
on $M^{(3)}_{ext}$ according to the maximum principle for subharmonic functions. In other words $\varphi$ is a geodesic map. 

Having once established that $\varphi$ is a geodesic map we can relate the scalar charges to the ''initial velocity"
of the geodesic $\psi$. Indeed we get

\begin{eqnarray}
q^a=\frac{1}{4\pi} \int_{S^2_\infty} D_i\varphi^a d\Sigma^i(h)=\frac{d\psi^a}{du}(0) \frac{1}{4\pi} \int_{S^2_\infty} D_iu d\Sigma^i(h)=
\frac{d\psi^a}{du}(0)M
\end{eqnarray} 
or equivalently $\frac{d\psi^a}{du}(0)=\frac{q^a}{M}$. The geodesic $\psi$, and as a consequence also the geodesic map ${\tilde \varphi}$, is uniquely specified by the asymptotic data $(\varphi^a_{\infty},q^a/M)$ which in turn means that for fixed $(\varphi^a_{\infty},q^a, M)$ there is only one possible solution to the scalar field equations, namely the geodesic map ${\tilde \varphi}$  with the asymptotic data $(\varphi^a_{\infty},q^a/M)$. This completes the proof of the theorem.  

\section{Some relations for the photon sphere geometric characteristics and the asymptotic charges}

In this section we shall derive some key relations among the geometrical characteristics of the photon sphere and the asymptotic charges.

Let us denote by  $p$ the metric induced on the photon surface  ${\cal P}$ and by ${\cal K}^{\cal P}$ its second fundamental form. A key result in the theory of the photon surfaces is that  ${\cal P}$  is  totally umbilic \cite{Claudel2001,Perlick2005}, i.e. 

\begin{eqnarray}\label{PEK}
{\cal K}^{\cal P}= \frac{H^{\cal P}}{3}p ,
\end{eqnarray}
where $H^{\cal P}$ is the mean curvature of ${\cal P}$. It is easy one to show that $H^{\cal P}$
is constant on  ${\cal P}$. Making use of the contracted Codazzi equation for $({\cal P}, p)\hookrightarrow (M^{(4)}, g^{(4)})$ we get

\begin{eqnarray}\label{CH}
\frac{2}{3}X(H^{\cal P})= Ric^{(4)}(X, n)=2 \gamma_{ab}(\varphi_0)X(\varphi^a)n(\varphi^b)=0 ,
\end{eqnarray}
where $n$ is the unit normal to ${\cal P}$ and $X$ is a vector field tangent to ${\cal P}$, i.e. $X\in \Gamma (T{\cal P})$.
In the last step we have taken into account that $\varphi$ is constant on ${\cal P}$. The equality  (\ref{CH}) shows that $H^{\cal P}$ is indeed constant.

For the second fundamental form ${\cal K}^{\Sigma}$ of  $(\Sigma,\sigma)\hookrightarrow (M^{(3)},g^{(3)})$ (with a unit normal $n$) we have
\begin{eqnarray}
{\cal K}^{\Sigma}(X, Y)= g^{(3)}(\nabla^{(3)}_{X}n, Y)= g^{(4)}(\nabla_{X}n, Y)= {\cal K}^{\cal P}(X,Y)= \frac{H^{\cal P}}{3} p(X,Y)= \frac{H^{\cal P}}{3} \sigma(X,Y) ,
\end{eqnarray}
where $X,Y \in \Gamma(T\Sigma)$. Therefore, we find that ${\cal K}^{\Sigma}=\frac{H^{\cal P}}{3}\sigma$ which also gives a simple relation between the mean curvatures $H^{\cal P}$
and $H^{\Sigma}$, namely $H^{\Sigma}=\frac{2}{3}H^{\cal P}$. Using this information in the contracted Codazzi equation we easily find that $Ric^{(3)}(X,n)=0$.

Now let us consider  $n(N)$ and show that it is a constant on $\Sigma$. For  an arbitrary $X\in\Gamma(T\Sigma)$ we have

\begin{eqnarray}
X(n(N))= X(n(N)) - (\nabla^{(3)}_{X}n)(N)= (\nabla^{(3)}\nabla^{(3)}N)(X,n)= \nonumber \\ N \left[Ric^{(3)}(X,n) - 2\gamma_{ab}(\varphi_0)X(\varphi^a)n(\varphi^b)\right]=0,
\end{eqnarray}
where we have taken into account the dimensionally reduced field equations (\ref{DRFE}) and the fact that $N$ and $\varphi$ are constant on $\Sigma$. Therefore $n(N)$  is indeed constant on $\Sigma$. As an immediate consequence we have that $\gamma_{ab}(\varphi_0)n(\varphi^a)n(\varphi^b)$ is also a constant on $\Sigma$ since $\varphi$ is a geodesic map.

From the Gauss equation for $(\Sigma, \sigma)\hookrightarrow ({\cal P},p)$, and  taking into account that the spacetime is static, it is  easy to show
that the Ricci scalar curvature $R^{\Sigma}$ of $\Sigma$ is given by

\begin{eqnarray}\label{RSCS1}
R^{\Sigma}= R^{\cal P}= \frac{2}{3} (H^{\cal P})^2 - 2\gamma_{ab}(\varphi_0)n(\varphi^a)n(\varphi^b) =
\frac{3}{2} (H^{\Sigma })^2 - 2\gamma_{ab}(\varphi_0)n(\varphi^a)n(\varphi^b)  .
\end{eqnarray}

Further the Gauss equation

\begin{eqnarray}
R^{(3)} - 2 Ric^{(3)}(n,n)= R^{\Sigma} - \left(Tr {\cal K}^{\Sigma}\right)^2 + Tr\left( {\cal K}^{\Sigma}\right)^2
\end{eqnarray}
for  $(\Sigma,\sigma)\hookrightarrow (M^{(3)},g^{(3)})$ gives

\begin{eqnarray}\label{Ric1}
R^{(3)} - 2 Ric^{(3)}(n,n)=R^{\Sigma} - \frac{1}{2} (H^{\Sigma})^2 .
\end{eqnarray}

In order to calculate $Ric^{(3)}(n,n)$ we can use the dimensionally reduced field equations (\ref{DRFE}) and

\begin{eqnarray}
\Delta^{(3)}N= \Delta^{(2)}N + \nabla^{(3)}\nabla^{(3)}N(n,n) + H^{\Sigma}n(N),
\end{eqnarray}
which leads to

\begin{eqnarray}\label{Ric2}
N_{0} Ric^{(3)}(n,n)= -H^{\Sigma} n(N) + 2N_{0}\gamma_{ab}(\varphi_0)n(\varphi^a)n(\varphi^b).
\end{eqnarray}

Using again the dimensionally reduced field equations it is not difficult one to show that $R^{(3)}=2\gamma_{ab}(\varphi_0)n(\varphi^a)n(\varphi^b)$ which combined with (\ref{Ric1}) and (\ref{Ric2}) gives

\begin{eqnarray}\label{RSCS2}
N_{0}R^{\Sigma}= 2H^{\Sigma}n(N) + \frac{1}{2} N_{0} (H^{\Sigma})^2 - 2N_{0}\gamma_{ab}(\varphi_0)n(\varphi^a)n(\varphi^b).
\end{eqnarray}

Integrating  (\ref{RSCS1}) and (\ref{RSCS2}) over $\Sigma$ and taking into account the Gauss-Bonnet theorem $\int_{\Sigma} R^{\Sigma}\sqrt{\sigma}d^2x=8\pi$ for the topological sphere $\Sigma$ we get the following relations

\begin{eqnarray}\label{IDFHN}
&& 1= \frac{3}{16\pi}(H^{\Sigma})^2 {\cal A}_{\Sigma} - \frac{1}{4\pi} \gamma_{ab}(\varphi_0)n(\varphi^a)n(\varphi^b) {\cal A}_{\Sigma}, \\
&& N_{0}= \frac{1}{4\pi} H^{\Sigma} n(N) {\cal A}_{\Sigma} +  \frac{1}{16\pi}N_{0} (H^{\Sigma})^2 {\cal A}_{\Sigma}  -  \frac{1}{4\pi} N_{0}\gamma_{ab}(\varphi_0)n(\varphi^a)n(\varphi^b) {\cal A}_{\Sigma}.\nonumber
\end{eqnarray}

These relations can be further simplified  by taking into account that $\varphi$ is a geodesic map, namely

\begin{eqnarray}\label{IDFHN}
&& 1= \frac{3}{16\pi}(H^{\Sigma})^2 {\cal A}_{\Sigma} - \frac{N^{-2}_0}{4\pi} \frac{Q^2}{M^2} (n(N))^2 {\cal A}_{\Sigma} , \\
&& N_{0}= \frac{1}{4\pi} H^{\Sigma} n(N) {\cal A}_{\Sigma} +  \frac{1}{16\pi}N_{0} (H^{\Sigma})^2 {\cal A}_{\Sigma}  -  \frac{N^{-1}_0}{4\pi} \frac{Q^2}{M^2} (n(N))^2 {\cal A}_{\Sigma},\nonumber
\end{eqnarray}
where $Q$ is the generalized scalar charge. Multiplying these equations by ${\cal A}_{\Sigma}$  and using that $n(N){\cal A}_{\Sigma}=4\pi M$  one finds

\begin{eqnarray}
&&M=\frac{1}{8\pi}N_{0}H^{\Sigma} A_{\Sigma}, \\
&&\frac{N^{2}_0 }{4\pi}A_{\Sigma}= 3M^2 -Q^2. 
\end{eqnarray} From the second relation we see that the presence of a photon sphere 
imposes a restriction on the generalized scalar charge, namely  $Q^2< 3 M^2$.

\section{Classification theorem}

The main result of the present paper is the following

\medskip
\noindent

{\bf Theorem} {\it Let us consider the Einstein-multiple-scalar field equations  with a target space which is a complete, simply connected Riemannian manifold with a non-positive sectional curvature. Then  there can be only one static and asymptotically flat spacetime $(M_{ext}^{(4)}, g^{(4)},\varphi)$,  satisfying the static Einstein-multiple-scalar field equations, possessing a photon sphere ${\cal P}\hookrightarrow M_{ext}^{(4)}$ as an inner boundary of $M_{ext}^{(4)}$, with a lapse function $N$ regularly foliating  $M_{ext}^{(4)}$ and given asymptotic data $(M, q^a, \varphi^a_{\infty})$ with a generalized scalar charge $Q$ satisfying $Q^2<3M^2$. The  spacetime is isometric to the Janis-Newman-Winicour solution with $\nu=\sqrt{1+ \frac{Q^2}{M^2}} $. }

\medskip
\noindent

{\bf Proof:} The fact that $\varphi$ is a geodesic map (i.e. $\varphi^a=\varphi^a(u)$) reduces the problem to the case with a single scalar field.
Then the proof of our classification theorem follows directly from the theorem proven in \cite{Yazadjiev2015}. As a consequence we have that the spacetime and the scalar fields are spherically symmetric.

In order to explicitly construct the unique solution with a  photon sphere and asymptoic data $(M, q^a, \varphi^{a}_\infty)$ we shall 
use the dymensonally reduced equations (\ref{DRFEH}). Using the fact that $\varphi$ is a geodesic map the equations (\ref{DRFEH}) reduce further to 

\begin{eqnarray}
&&R(h)_{ij}= 2D_i{\hat u}D_j{\hat u}, \\
&&D_iD^i {\hat u}= 0 
\end{eqnarray} 
where ${\hat u}=u/\nu$ with $\nu^{-1}=\sqrt{1 + \frac{Q^2}{M^2}}$. We notice that these equations are in fact  the static vacuum Einstein equations written in terms of the metric $h_{ij}$ with an effective gravitational potential  ${\hat u}$ having the asymptotic 
${\hat u}\approx  \frac{M}{\nu r} + O(r^{-2})$. Knowing that the Schwarzschild solution is the only static and spherically symmetric solution of the vacuum Einstein equations we find

\begin{eqnarray}
e^{2{\hat u}}=1 - \frac{2M}{\nu r}, \;\;\;\; h_{ij}dx^{i}dx^{j}= \left(1 - \frac{2M}{\nu r}\right)\left[\frac{dr^2}{1- \frac{2M}{\nu r}} + r^2 d\Omega^2_{S^2}\right] ,
\end{eqnarray}
where $d\Omega^2_{S^2}$ is the standard metric on the unit 2-dimensional sphere. For the original lapse function $N=e^u=e^{\nu {\hat u}}$   and the 3-metric $g^{(3)}_{ij}$ we  get

\begin{eqnarray}
&&N^2= \left(1 - \frac{2M}{\nu r}\right)^{\nu}, \\
&&g^{(3)}_{ij}dx^idx^j= N^{\,-2} h_{ij}dx^i dx^j=  \left(1 - \frac{2M}{\nu r}\right)^{-\nu} dr^2  +  \left(1 - \frac{2M}{\nu r}\right)^{1-\nu} r^2 d\Omega^2_{S^2}.
\end{eqnarray}

Having the explicit expressions for the lapse function and the 3-metric we can write the desired  4-dimensional metric

\begin{eqnarray}
&&ds^2=g^{(4)}_{\mu\nu}dx^{\mu}dx^{\nu}= - \left(1 - \frac{2M}{\nu r}\right)^{\nu} dt^2 + \left(1 - \frac{2M}{\nu r}\right)^{-\nu} dr^2  +  \left(1 - \frac{2M}{\nu r}\right)^{1-\nu} r^2 d\Omega^2_{S^2}, \nonumber\\
\end{eqnarray}
which is just the Janis-Newman-Winicour solution specified only by the mass $M$ and the generalized scalar charge $Q$ with $Q^2<3M^2$. 
The explicit form of the scalar fields is given by the explicit form of  the geodesics with asymptotic data $(q^a/M,\varphi^a_\infty)$ with an affine parameter $u= \frac{\nu}{2} \left(1 - \frac{2M}{\nu r}\right)$. This completes the proof of the theorem.

\section{Discussion}

In the present paper we have proven that the static and asymptotically flat solutions to the  Einstein-multiple-scalar field theory  possessing a photon sphere are uniquely specified by their asymptotic data, i.e. by their   mass, scalar charges and asymptotic values of the scalar fields. The proof is under certain assumptions.  The first assumption is that the lapse function $N$ regularly  foliates the exterior spacetime. This assumption can be easily relaxed -- in fact the proof in \cite{Yazadjiev2015} does not need this assumption. However, the relaxation of this assumption leads to nothing essential from a physical point of view. The second  assumption is that the target space is  a complete, simply connected Riemannian manifold with a non-positive sectional curvature.  The natural question is whether the proof can be extended to the case of target spaces with positive curvature. Our preliminary studies show that the proof can be extended  when the positive curvature is bounded from above, however, with some restrictive  conditions on the scalar map $\varphi$ or the lapse function $N$ itself. At this stage it is not clear how one can encode these restrictive conditions on $\varphi$  or $N$ in the asymptotic data only. For completeness let us present our conjecture. Let us assume that the target space sectional curvature  $K$  satisfies $K\le K_{+} $ where $K_{+}\ge 0$. Then  we conjecture that our uniqueness result holds when, in addition to  some other mild assumptions, the lapse function on the photon sphere obeys the inequality 
\begin{eqnarray}
N^2_0	< e^{\frac{\pi M}{Q\sqrt{K_+}}}.?
\end{eqnarray}
We hope to return to this problem in a future publication.

\medskip
\noindent

\noindent {\bf Acknowledgements:} 
The partial support by the Bulgarian NSF Grant KP-06-RUSSIA/13 and the Networking support by the COST Action CA16104 are also gratefully acknowledged.

\end{document}